\documentstyle[prb,aps,twocolumn,psfig]{revtex}
\begin{document}
\draft
\title{Ground state properties of heavy alkali halides}
\author{
  Klaus Doll}
\address{
     Max-Planck-Institut f\"ur Physik komplexer Systeme,
          N\"othnitzer Strasse 38, D-01187 Dresden, Germany }
\author{Hermann Stoll}

\address{
Institut f\"ur Theoretische Chemie, Pfaffenwaldring 55, 
Universit\"at Stuttgart, D-70550 Stuttgart,
Germany
}

\maketitle

\begin{abstract}
We extend previous work on 
alkali halides by calculations for the heavy-atom species 
RbF, RbCl, LiBr, NaBr, KBr, RbBr, LiI, NaI, KI, and RbI. 
Relativistic effects are included by means of energy-consistent 
pseudopotentials,
correlations are treated at the coupled-cluster
level. 
A striking deficiency of the 
Hartree-Fock approach are
lattice constants deviating by up to 7.5 \% from experimental values
which is reduced to a maximum error of 2.4  \% 
by taking into account electron correlation. 
Besides, we provide {\em ab-initio} data for 
in-crystal polarizabilities and van der Waals coefficients. 
\end{abstract}

\pacs{ }

\narrowtext

Calculations for solids 
based on the {\em exact} Hartree-Fock (HF) exchange have 
been widely used for determining total energies,
lattice constants, elastic properties, phase transitions, equations of 
state, and also magnetic properties \cite{CRYSTALbuch,PisaniBuch}. 
In Ref. \onlinecite{Stoll}, it was shown how a correlation treatment 
with quantum-chemical methods could be performed for semiconductors.
We extended this to ionic systems \cite{DDFS} and recently presented results
for six light alkali halide systems \cite{Alkali}; 
we now turn to their
heavier counterparts, where correlation effects are stronger and 
relativistic effects should be included. We restrict
 ourselves to a comparison of different systems with NaCl structure. 
The method has been explained in Ref. \onlinecite{Alkali}.

\section{Hartree-Fock calculation}
We performed 
self-consistent-field (SCF) calculations 
with the program package 
{\sc{Crystal 95}} (Ref. \onlinecite{Manual}).
For Li, Na, K, F and Cl, 
we used the basis sets given in Ref. \onlinecite{Prencipe}.
The most diffuse $sp$-exponents were reoptimized 
(Table \ref{CRYSTALbasis}). For the free metal atoms one more 
diffuse $sp$ shell was added 
($sp$-exponents are 0.048 for Li, 0.040 for Na, and 0.028 for K).
For Rb, Br, and I, we used relativistic pseudopotentials (for details cf.\
section \ref{PPsection})
 in combination with the
corresponding basis sets, again reoptimizing
diffuse exponents (Table \ref{CRYSTALbasis}).
For Rb, the first three $s$ and $p$ exponents were kept contracted from the 
atom-optimized basis set,
 and two $sp$ exponents  optimized in calculations for
the solid were added, yielding a [3s3p] basis set after these steps.
For Br and I, the two tight
atomic $s$ and $p$ exponents were kept fixed, and two diffuse $s$ and $p$ 
exponents
were optimized which resulted in a [4s4p] basis set.
We optimized one
$d$ function for the heavier alkali metal ions (Table \ref{CRYSTALbasis}).
This leads to a tiny increase of the cohesive energy and a reduction of
the bond length of $\lesssim$ 0.5\%. Adding $d$-functions to the anions 
virtually does not change the results (see also Ref. \onlinecite{Prencipe}).
Cohesive
energies with respect to neutral atoms are strongly underestimated at the 
SCF level, by up to 29\%.
Lattice constants are too large by up to 7.5 \% and, as a consequence, 
bulk moduli are underestimated by up to  36\% (see also  earlier calculations
\cite{PrencipeThesis,PisaniBuch}).

\section{Correlation calculation}

\label{PPsection}
We performed correlation calculations using the coupled cluster approach
with single and double substitutions (CCSD) and with the inclusion of
triples within the framework of perturbation theory (CCSD(T)\cite{CCSDT})
using the program package {\sc Molpro}
\cite{KnowlesWerner,MOLPROpapers}. 
For Li, Na, K, F, and Cl, we used the same basis sets as
in Ref. \onlinecite{Alkali}. For Rb, we applied the scalar-relativistic
energy-consistent small-core (9-valence-electron)
pseudopotential \cite{Leininger} together with the corresponding
atom-optimized basis set
(uncontracted),  and
augmented the latter with 5 $d$ functions and 3 $f$ functions,
resulting in a $[7s6p5d3f]$ basis set 
($d$ exponents  are 0.801 900, 0.297 000, 0.110 000,
0.040 741, 0.015 089, $f$ exponents are 1.320 000, 0.600 000, 0.272 727).
For Br and I, 7-valence-electron pseudopotentials, again of
the scalar-relativistic energy-consistent variety, were used \cite {Bergner}.
$s$ and $p$ functions were left uncontracted, three $d$ and
two $f$ functions were added, so that the final basis sets 
are $[4s5p3d2f]$ 
(Br $d$ exponents are 0.604 185, 0.271 744, and 
0.095, $f$ exponents are 0.58 and 0.26; I $d$ exponents are
0.367 696, 0.183 848, and 0.07, $f$ exponents are 0.43 and 0.19).

In Table \ref{IPEA}, we show atomic ionization potentials (IP) and electron 
affinities (EA).
The correlation contribution to the EA decreases
from F to I with increasing size of the atoms:
the additional electron completes the shell and correlations are more 
important in shells with smaller radii.
In the metal atoms, the correlation contribution
to the IP (or the EA of the corresponding cations)
increases from Li to Rb.
The additional electron is the only electron in the 
open shell and the main correlation effect is a dynamical core polarization.
The correlation energy is  $-\frac{1}{2}\alpha \vec E^2$
($\alpha$  is the core polarizability, $\vec E$ 
is the instantaneous
field of the valence electron at the site of the core
\cite{Mueller}).
The higher polarizabilities of the heavier ions 
lead to larger absolute correlation energies.

\section{Results for the solid and discussion}

In the solid, 
the correlation energies of the cations  are nearly the same as for free
ions. The correlation energies of the diffuse anions
become significantly smaller in magnitude 
(cf. Tables \ref{inkrubidium},\ref{inkbromides},\ref{inkiodides}).
This originates from the upward energy shift of
excited-state determinants when the anions are compressed.
The difference in correlation energies
$|\epsilon_{corr}({\rm free \mbox{ } anion})
-\epsilon_{corr}{\rm(embedded \mbox{ } anion)}|$ 
decreases
from F to I: since the size
of the valence shell is smaller for atoms with low main quantum number, 
the concomitant
correlation energy is larger and the compression due to the neighboring ions
leads to a stronger change of the correlation energy.

Two-body contributions are shown in Tables 
\ref{inkrubidium},\ref{inkbromides},\ref{inkiodides}. 
Their magnitude is essentially determined by the distance and the 
polarizabilities. 
We evaluated two-body increments typically up to third
nearest neighbors for halide-halide and metal-halide pairs, metal-metal 
increments for nearest neighbors only.
We used the two-body increments for next neighbors to
calculate van der Waals $C_6$ coefficients ($E_{vdW}=-\frac{C_6}{r^6}$). 
Polarizabilities increase with
increasing atomic radii and electron number from Li to Rb and from F to I,
but are roughly constant for the same ion in different halides
(cf.\  Table \ref{vdWTabelle}).
Evaluating London's formula gives a 
qualitative explanation of the relative strength of the van der Waals
interaction in different systems.

Including electron
correlation with CCSD(T), we obtain nice agreement with experiment with a 
deviation of 7 \% for cohesive energies or 5\% for lattice energies in the
worst case (Table \ref{Summen}). 
For lattice constants, the remaining largest deviation from experiment at the
CCSD(T) level is 2.4 \%. 
Bulk moduli are consistently in better agreement with experiment 
after
including correlations, at the CCSD(T) level the maximum deviation is 15 \%
or 6 \% on average.
We are aware of one density functional calculation for
these systems \cite{Cortona}. 
Although the local density approximation was applied, i.e. no gradient
correction, the 
agreement with experiment is not worse than with our approach.
Still, our approach has the advantage of being free of parameters
and calculation of EAs or IPs
does not pose problems.
Moreover, a detailed interpretation of correlation effects is possible.
However, for calculating the correlation energy at one lattice constant, 
$\lesssim 10$  hours of CPU time on workstations are necessary 
(when three-body corrections are neglected which are
the most expensive ones),
whereas a density functional calculation will be much faster (analogously, the 
{\sc Crystal}-SCF calculation takes only a few minutes).

\onecolumn
\begin{table}
\begin{center}
\caption{\label{CRYSTALbasis}Diffuse exponents of Gaussian basis sets, 
and polarization functions, for the CRYSTAL calculations}
\vspace{5mm}
\begin{tabular}{|c|cc|}  
System & cation & anion \\ & & \\
RbF    &  $sp$ 0.336, 0.136, $d$ 0.38 &  $sp$ 0.429, 0.133\\
RbCl   & $sp$ 0.334, 0.140, $d$ 0.30 &  $sp$ 0.322, 0.114, $d$ 0.50 \\
LiBr   &  $sp$ 0.513 &  $s$ 0.400, 0.145, $p$ 0.324, 0.107 \\
NaBr   &  $sp$ 0.538, 0.198 &  $s$ 0.389, 0.141, $p$ 0.315, 0.107 \\
KBr    &  $sp$ 0.394, 0.219, $d$ 0.42 &  $s$ 0.392, 0.140, $p$ 0.326, 0.108
\\
RbBr   &  $sp$ 0.336, 0.141, $d$ 0.31 &  $s$ 0.393, 0.140, $p$ 0.323, 0.105 \\
LiI  &  $sp$ 0.511 &  $s$ 0.327, 0.122, $p$ 0.322, 0.101 \\
NaI  &  $sp$ 0.533, 0.176 &  $s$ 0.323, 0.124, $p$ 0.321, 0.104 \\
KI  & $sp$ 0.397, 0.219, $d$ 0.45 &  $s$ 0.324, 0.121, $p$ 0.321, 0.101 \\
RbI &  $sp$ 0.333, 0.141, $d$ 0.31 &  $s$ 0.322, 0.120, $p$ 0.322, 0.099 \\
\end{tabular}
\end{center}
\end{table}

\begin{table}
\begin{center}
\caption{\label{IPEA}Electron affinities (EA) for halogen atoms
and ionization potentials (IP) for alkali atoms,
in Hartree units. The second column gives the value for EA or IP at the
Hartree-Fock level, the fourth and sixth at correlated levels. The third
and fifth column provide separate correlation contributions to EA and IP at
CCSD and CCSD(T) level, respectively. Finally, the last column shows the
experimental values. Since our calculated values for the neutral halogen
atoms are without spin-orbit coupling, i.e. averaged over the $^3P_{1/2}$ and
$^3P_{3/2}$ states, we give in
brackets experimental results after averaging over these two states.}
\vspace{5mm}
\begin{tabular}{|ccccccc|}
System & HF=EA/IP (HF) 
 & CCSD & EA/IP (CCSD) & CCSD(T) & EA/IP (CCSD(T)) &  exp. EA/IP 
\cite{CRC,Moore} \\ \hline
  F $\rightarrow$ F$^{-}$ & 0.050704 & 0.065415 & 0.116119 & 0.071218 &
  0.121922 & 0.12499 (0.12560) \\
  Cl $\rightarrow$ Cl$^{-}$ & 0.095053 & 0.030997 & 0.126051 &
 0.034132 &  0.129186 & 0.13276 (0.13410) \\
  Br $\rightarrow$ Br$^{-}$ & 0.093216 & 0.026662 & 0.119877 & 0.029237 &
 0.122452 & 0.12361 (0.12921) \\
  I $\rightarrow$ I$^{-}$ & 0.092016 & 0.021998 & 0.114014 & 0.024216 & 
 0.116232 & 0.11242 (0.12397) \\
  Li $\rightarrow$ Li$^{+}$ & 0.196311 & 0.000994 & 0.197305 & 0.001023 & 
 0.197334 & 0.19814 \\
  Na $\rightarrow$ Na$^{+}$ & 0.181948 & 0.005904 & 0.187852 & 0.006156 & 
 0.188104 & 0.18886 \\
  K  $\rightarrow$ K$^{+}$ & 0.146785 & 0.009586 & 0.156371 & 0.010446 & 
 0.157231 & 0.15952 \\
  Rb $\rightarrow$ Rb$^+$ & 0.139571 & 0.011481 & 0.151052 & 0.012392 & 
 0.151963 & 0.15351 \\
\end{tabular}
\end{center}
\end{table}

\begin{table}
\begin{center}
\caption{\label{inkrubidium}Local correlation energies 
per primitive unit cell (in Hartree)
for RbF ( at a lattice constant of 5.65 \AA) and RbCl (6.65 \AA)}
\vspace{5mm}
\begin{tabular}{|c|cc|cc|} 
 & \multicolumn{2}{c|}{RbF} & \multicolumn{2}{c|}{RbCl} \\
 & \multicolumn{1}{c}{CCSD}  & \multicolumn{1}{c|}{CCSD(T)} 
 & \multicolumn{1}{c}{CCSD}  & \multicolumn{1}{c|}{CCSD(T)} \\
\hline
 {free Rb$^{+}$} $\rightarrow$ \rm {embedded Rb$^{+}$} & -0.000015 & -0.000018
  & -0.000005 & -0.000006  \\
 {free X$^{-}$}  $\rightarrow$ \rm {embedded X$^{-}$} &   +0.010103 & 
+0.012906 & +0.002433 & +0.003435  \\
 {sum of X-X increments} & -0.000753 & -0.000879 & -0.002465 & -0.002915 \\
 {sum of Rb-X increments}  & -0.016480 & -0.018827 &  -0.016480 & -0.018736 \\ 
 {sum of Rb-Rb increments} & -0.002718 & -0.003021 & -0.000802 & -0.000893 \\ 
{ } & & & &  \\ 
 {total sum  } & -0.009863 & -0.009839 & -0.017319 & -0.019115 \\
\end{tabular}
\end{center}
\end{table}

\begin{table}
\begin{center}
\caption{\label{inkbromides}Local 
correlation energies per primitive unit cell (in Hartree)
for LiBr (at a lattice constant of 5.50 \AA), NaBr (5.98 \AA), KBr 
(6.70 \AA), and RbBr (7.00 \AA).}
\vspace{5mm}
\begin{tabular}{|c|cc|cc|cc|cc|} 
 & \multicolumn{2}{c|}{LiBr} & \multicolumn{2}{c|}{NaBr}
& \multicolumn{2}{c|}{KBr} & \multicolumn{2}{c|}{RbBr} \\
 & \multicolumn{1}{c}{CCSD}  & \multicolumn{1}{c|}{CCSD(T)} 
 & \multicolumn{1}{c}{CCSD}  & \multicolumn{1}{c|}{CCSD(T)} 
 & \multicolumn{1}{c}{CCSD}  & \multicolumn{1}{c|}{CCSD(T)} 
 & \multicolumn{1}{c}{CCSD}  & \multicolumn{1}{c|}{CCSD(T)} \\
\hline
 {free M$^{+}$} $\rightarrow$ \rm {embedded M$^{+}$} & -0.000011 &
 -0.000011 & -0.000086 & -0.000093 & -0.000003 & -0.000004 &
-0.000004 & -0.000004\\
 {free Br$^{-}$}  $\rightarrow$ \rm {embedded Br$^{-}$} &   +0.001802 & 
+0.002489 & +0.001778 & +0.002455   &  +0.001682  & +0.002353 & +0.001716 & 
+0.002422 \\
 {sum of Br-Br increments} & -0.015871 & -0.018428 & -0.009360  & -0.010917 
& -0.004218  & -0.004933 & -0.002957 & -0.003452 \\
 {sum of M-Br increments}  & -0.002429 & -0.002599 & -0.006814 & -0.007449
 & -0.013715  & -0.015675 & -0.015936 & -0.018034 \\ 
 {sum of M-M increments}   & \multicolumn{2}{c|}{absolute value $<$10$^{-6}$}
 & -0.000037 & -0.000040 & -0.000283  & -0.000320 & -0.000569 & -0.000633 \\
{ } & & & & & & & & \\ 
{total sum } & -0.016509 & -0.018549 & -0.014519 & -0.016044 & -0.016537  &  
-0.018579 & -0.017750 & -0.019701 \\
\end{tabular}
\end{center}
\end{table}

\begin{table}
\begin{center}
\caption{\label{inkiodides}Local 
correlation energies per primitive unit cell (in Hartree)
for LiI (at a lattice constant of 6.05 \AA), NaI (6.50 \AA), KI 
(7.20 \AA), and RbI (7.45 \AA).}
\vspace{5mm}
\begin{tabular}{|c|cc|cc|cc|cc|} 
 & \multicolumn{2}{c|}{LiI} & \multicolumn{2}{c|}{NaI}
& \multicolumn{2}{c|}{KI} & \multicolumn{2}{c|}{RbI} \\
 & \multicolumn{1}{c}{CCSD}  & \multicolumn{1}{c|}{CCSD(T)} 
 & \multicolumn{1}{c}{CCSD}  & \multicolumn{1}{c|}{CCSD(T)} 
 & \multicolumn{1}{c}{CCSD}  & \multicolumn{1}{c|}{CCSD(T)} 
 & \multicolumn{1}{c}{CCSD}  & \multicolumn{1}{c|}{CCSD(T)} \\
\hline
 {free M$^{+}$} $\rightarrow$ \rm {embedded M$^{+}$} & -0.000007 &
 -0.000007 & -0.000070 & -0.000076 & -0.000002 & -0.000003 &
 -0.000002 & -0.000003 \\
 {free I$^{-}$}  $\rightarrow$ \rm {embedded I$^{-}$} &   +0.001007 & 
+0.001470 & +0.001038 & +0.001503   &  +0.001031  & +0.001508 & +0.001088 & 
+0.001604 \\
 {sum of I-I increments} & -0.018703 & -0.021757 & -0.011833  &  -0.013815
& -0.005666  & -0.006632  & -0.004139 &  -0.004835 \\
 {sum of M-I increments}  & -0.002153 & -0.002304 & -0.006200 & -0.006776
 & -0.012961 & -0.014807 & -0.015830 & -0.017900 \\ 
 {sum of M-M increments}   & \multicolumn{2}{c|}{absolute value $<$10$^{-6}$}
 & -0.000022 & -0.000024 & -0.000178  & -0.000202 & -0.000377 & -0.000420 \\
{ } & & & & & & & & \\ 
{total sum } & -0.019856 & -0.022598 & -0.017087 & -0.019188 & -0.017776  &  
-0.020136 & -0.019260 & -0.021554 \\
\end{tabular}
\end{center}
\end{table}
\twocolumn

\onecolumn
\begin{table}
\begin{center}
\caption{\label{vdWTabelle}Comparison of CCSD two-body increments 
$\Delta E$ between next
neighbors (weight factors for the solid not included). All results are
given in atomic units (except for the lattice constant in column 2). 
$r$ is the distance
between the respective ions in bohr. IP$_{cat}$ and IP$_{an}$ are in-crystal
ionization
potentials for the respective ions, $\alpha_{cat}$ and $\alpha_{an}$ the
corresponding polarizabilities. In the last column, results from CCSD
calculations are divided by those obtained from a simple estimate using
the London formula.}
\vspace{5mm}
\begin{tabular}{|ccccccccc|} 
 System & lattice & \multicolumn{1}{c}{$\Delta E$}  &
 $-C_{6}=\Delta E$ $\times$ $r^6$ & IP$_{cat}$ & IP$_{an}$ &
$\alpha_{cat}$ & $\alpha_{an}$ & $ -\frac{2}{3}\frac{r^6}{\alpha_{1}
\alpha_{2}}\frac{IP_1 + IP_2}{IP_1 IP_2}$ $\times$ $\Delta E$\\
& constant $a$ in \AA &  & & & & & & \\
\hline
Li-Li (LiF) & 3.99 & $\sim$ -0.000003  & $\sim$ -0.07 & 2.3 &  & 0.19 &  & 
$\sim$ 1.1 \\
F-F (RbF) & 5.65 & -0.000111 & -20.6 & 0.67 & 0.40 & 8.9 & 5.1 & 2.6 \\
Rb-F (RbF) & 5.65 & -0.002625 & -60.8 & & & & & 3.6 \\
Rb-Rb (RbF) & 5.65 & -0.000411 & -76.1 & & & & & 1.9 \\
Cl-Cl (RbCl) & 6.65 & -0.000370 & -182 & 0.72 & 0.37 & 8.9 & 17.8 & 2.1 \\
Rb-Cl (RbCl) & 6.65 & -0.002626 & -162 & & & & & 2.8 \\
Rb-Rb (RbCl) & 6.65 & -0.000134  & -66.0 &  &  &  &  & 1.5 \\
Br-Br (LiBr) & 5.50 & -0.002423 & -382 & 2.4 & 0.43 & 0.19 & 23.1 & 2.2\\
Li-Br (LiBr) & 5.50 & -0.000405 & -7.98  & & & & & 3.3 \\
Br-Br (NaBr) & 5.98 & -0.001431 & -373 & 1.4 & 0.40 & 0.97 & 23.3 & 2.3 \\
Na-Br (NaBr) & 5.98 & -0.001090 & -35.5 & & & & & 3.4 \\ 
Na-Na (NaBr) & 5.98 & -0.000006 & -1.56 & & & & & 1.6 \\
Br-Br (KBr) & 6.70 & -0.000641 & -330 & 0.87 & 0.37 & 5.4 & 23.1 & 2.2 \\
K-Br (KBr) & 6.70 & -0.002193 & -141 & & & & & 2.9 \\
K-K (KBr) & 6.70 & -0.000047 & -24.2 & & & & & 1.3 \\
Br-Br (RbBr) & 7.00 & -0.000447 & -299 & 0.74 & 0.36 & 8.9 & 22.4 & 2.2 \\
Rb-Br (RbBr) & 7.00 & -0.002545 & -213 &  & & & & -2.9 \\
Rb-Rb (RbBr) & 7.00 & -0.000095 & -63.6 & & & & & 1.4\\
I-I (LiI) & 6.05 & -0.002863 & -799 & 2.4 & 0.40 & 0.19 & 35.0 & 2.2 \\
Li-I (LiI) & 6.05 & -0.000359 & -12.5 & & & & & 3.7 \\
I-I (NaI) & 6.50 & -0.001814 & -779 & 1.4 & 0.37 & 0.97 & 35.0 & 2.3 \\
Na-I (NaI) & 6.50 & -0.000995 & -53.4 & & & & & 3.6 \\
Na-Na (NaI) & 6.50 & $\sim$ -0.000004 & $\sim$ -1.72 & & & & & $\sim$ 1.7 \\
I-I (KI) & 7.20 & -0.000865 & -686 & 0.89 & 0.35 & 5.4 & 34.4 & 2.2 \\
K-I (KI) & 7.20 & -0.002078 & -206 & & & & & 2.9 \\
K-K (KI) & 7.20 & -0.000030 & -23.8 & & & & & 1.2 \\
I-I (RbI) & 7.45 & -0.000630 & -613 & 0.75 & 0.33 & 8.9 & 32.7 & 2.3 \\
Rb-I (RbI) & 7.45 & -0.002535 & -308 & & & & & 3.1 \\
Rb-Rb (RbI) & 7.45  & -0.000063 & -61.3 & & & & & 1.4 \\
\end{tabular}
\end{center}
\end{table}
\twocolumn

\onecolumn
\newpage
\begin{table}
\begin{center}
\caption{\label{Summen}Hartree-Fock (HF) 
and correlated results (CCSD, CCSD(T)), 
in comparison to density-functional (DFT) and experimental values, 
for the solids. Cohesive energies $E$ (with respect to neutral atoms) 
and lattice energies $E_{lat}$ (with respect to free ions) are
given in Hartree units (taken from Ref. \protect{\onlinecite{CRC}}), 
lattice constants $a$ in \AA \mbox{} (Ref. \protect{\onlinecite{Landolt}})
and bulk moduli $B$ in GPa. Zero point energies have been 
estimated with a Debye approximation (Debye temperatures taken from Ref.
\protect{\onlinecite{Debye}}) and added to the experimental cohesive
energies. 
Results for experimental cohesive energies averaged over $^3P_{1/2}$ and 
$^3P_{3/2}$ states of the neutral halogen atoms are given in brackets.
The experimental bulk moduli are at 4.2 K and 
have been taken from Ref. \protect{\onlinecite{Cortona}}
and references therein, except for KBr and LiI where data 
from Refs. \protect{\onlinecite{Haussuehl}} and 
\protect{\onlinecite{McLean}}, respectively,
have been extrapolated to zero temperature with the temperature
dependence taken from these references.}
\begin{tabular}{|cccccc|}
 & HF & CCSD & CCSD(T) & DFT (Ref. \onlinecite{Cortona})
& expt.
\\ \hline
RbF &  &  &  & & \\
$E_{lat}$ & 0.2839 & 0.2925 & 0.2925 &  & 0.304 \\ 
$E$ & 0.1946 & 0.2572 & 0.2620 & 0.279  & 0.276 (0.277) \\
$a$ & 5.819 & 5.671 & 5.657 & 5.73 & 5.588 \\
$B$ & 24.9 & 31.1  & 31.5 & 26.1  & 30.1 \\
RbCl & & & & & \\
$E_{lat}$ & 0.2407  & 0.2561  & 0.2579 & & 0.265 \\
$E$ & 0.1963 & 0.2312 & 0.2352 & 0.232 & 0.244 (0.245) \\
$a$ & 6.912 & 6.659 & 6.635 & 6.57 & 6.532\\
$B$ & 13.4 & 17.4 & 18.2 & 16.6 & 18.7 \\
LiBr & & & & & \\
$E_{lat}$ & 0.2863 & 0.3014 & 0.3034 & & 0.315 \\
$E$ &  0.1852 & 0.2260 & 0.2306 & 0.235 & 0.240 (0.246) \\
$a$  & 5.690 & 5.512 & 5.492 & 5.44 & 5.459 \\
$B$ & 24.6 & 27.5 & 27.5 & 28.5 & 26.3 \\
NaBr & & & & & \\
$E_{lat}$ & 0.2656 & 0.2790 & 0.2805 & & 0.288 \\
$E$ & 0.1777 & 0.2118 & 0.2156 & 0.206 & 0.223 (0.229) \\
$a$  & 6.181 & 5.984 & 5.966 & 6.10 & 5.926 \\
$B$ & 15.4 & 21.6 & 22.2 & 18.6 & 22.6 \\
KBr & & & & & \\
$E_{lat}$ & 0.2353 & 0.2504 & 0.2525 & & 0.264 \\
$E$ & 0.1849 & 0.2170 & 0.2208 & 0.220 & 0.228 (0.234) \\
$a$  & 6.939 & 6.685  & 6.654 & 6.57 & 6.541 \\
$B$ & 12.6 & 16.0 & 16.4 & 16.3 & 17.9 \\
RbBr &  & & & & \\
$E_{lat}$ & 0.2274 & 0.2434 & 0.2454 & & 0.255 \\
$E$ & 0.1812 & 0.2123 & 0.2160 & 0.213 & 0.225 (0.231) \\
$a$  & 7.269 & 6.996 & 6.967 & 6.87 & 6.822 \\
$B$ & 10.8 & 14.8 & 15.0 & 14.1 & 16.0 \\
LiI & & & & & \\
$E_{lat}$ & 0.2594 & 0.2775 & 0.2803 & & 0.293 \\
$E$ & 0.1572 & 0.1962 & 0.2012 & 0.202 & 0.207 (0.219)\\
$a$ & 6.299 & 6.034 & 6.002 & 5.91 & 5.946 \\
$B$ & 15.3 & 20.5 & 21.6 & 22.0 & 24.0 \\
NaI & & & & & \\
$E_{lat}$ & 0.2443 & 0.2596 & 0.2616 & & 0.269 \\
$E$ & 0.1551 & 0.1865  & 0.1906  & 0.173 & 0.193 (0.205)\\
$a$ & 6.756 & 6.525 & 6.499 & 6.58 & 6.409 \\
$B$ & 13.7 & 18.1 & 18.6 & 14.9 & 17.9 \\
KI & & & & & \\
$E_{lat}$ & 0.2181 & 0.2342 & 0.2366 & & 0.248 \\
$E$ & 0.1665 & 0.1949 & 0.1987 & 0.191 & 0.201 (0.213)\\
$a$  & 7.473 & 7.182 & 7.143 & 7.01 & 6.994 \\
$B$ & 10.5 & 12.9 & 13.1 & 13.8  & 12.7 \\
RbI & & & & & \\
$E_{lat}$ & 0.2116 & 0.2283 & 0.2306 & & 0.241 \\
$E$ & 0.1642 & 0.1914 & 0.1950 & 0.187 & 0.200 (0.212) \\
$a$  & 7.805 & 7.476 & 7.436 & 7.31 &  7.263 \\
$B$ & 8.6 & 11.0 & 11.2 & 12.2 & 13.1 \\
\end{tabular}
\end{center}
\end{table}


\begin{references}
\bibitem{CCSDT} After localization, we performed a rotation within the
group of orbitals to be correlated. 
Pseudo-canonical orbitals are obtained and  
the Fock matrix becomes diagonal in that block. This way, we have 
overcome the problem that the triples within 
CCSD(T) are only exact for orbitals for which the Fock matrix is
diagonal (see also M. J. O. Deegan and P. J. Knowles, Chem. Phys. Lett. 
{\bf 227}, 321 (1994)).
\end{references}
\end{document}